\documentclass[a4paper,twocolumn,showpacs,superscriptaddress,prl]{revtex4}

\usepackage[english]{babel}
\usepackage{xspace}

\usepackage{graphicx}
\usepackage{epsfig}
\usepackage{wrapfig}
\usepackage{dcolumn}
\usepackage{bm}
\newcounter{subfig}

\begin{document}
\title{Equilibrium shape and dislocation nucleation in strained epitaxial
nanoislands}

\author{J. Jalkanen}
\affiliation{Laboratory of Physics, P.O. Box 1100, Helsinki
University of Technology, FIN--02015 HUT, Espoo, Finland}

\author{O. Trushin}
\affiliation{Institute of Microelectronics and Informatics,
Academy of Sciences of Russia, Yaroslavl 150007, Russia}

\author{E. Granato}
\affiliation{Laborat\'orio Associado de Sensores e Materiais,
Instituto Nacional de Pesquisas Espaciais, 12245--970 S\~ao Jos\'e
dos Campos, SP Brasil}

\author{S. C. Ying}
\affiliation{Department of Physics, P.O. Box 1843, Brown
University, Providence, RI 02912--1843}

\author{T. Ala-Nissila}
\affiliation{Laboratory of Physics, P.O. Box 1100, Helsinki
University of Technology, FIN--02015 HUT, Espoo, Finland}
\affiliation{Department of Physics, P.O. Box 1843, Brown
University, Providence, RI 02912--1843}


\date{June 15, 2005}

\begin{abstract}
We study numerically the equilibrium shapes, shape transitions and
dislocation nucleation of small strained epitaxial islands with a
two-dimensional atomistic model, using simple interatomic pair
potentials. We first map out the phase diagram for the equilibrium
island shapes as a function of island size (up to $N=105$ atoms)
and lattice misfit with the substrate and show that nanoscopic
islands have four generic equilibrium shapes, in contrast with
predictions from the continuum theory of elasticity. For
increasing substrate-adsorbate attraction, we find islands that
form on top of a finite wetting layer as observed in
Stranski-Krastanow growth. We also investigate energy barriers and
transition paths for transitions between different shapes of the
islands and for dislocation nucleation in initially coherent
islands. In particular, we find that dislocations nucleate
spontaneously at the edges of the adsorbate-substrate interface
above a critical size or lattice misfit.
\end{abstract}


\maketitle

The shape and size of islands resulting from a  growth process has
been a subject of intense experimental and theoretical studies
\cite{politi00,daruka99,daruka02,wang,zangwill,uemura,%
johnson,spencer01,tersoff84,Budiman,chen}. There are still
uncertainties as to whether the observed shape and size of islands
corresponds to thermodynamic equilibrium state of minimum free
energy, or whether they are limited by kinetic effects. In an
equilibrium theory, the optimal size and shape result from a
delicate balance of energy lowering through strain relaxation and
energy cost through extra surface energy. Earlier works on
equilibrium shape of coherent islands have used simple predefined
faceted shapes in analytical calculations based on continuum
elasticity theory \cite{daruka99,daruka02}. The resulting
equilibrium shapes of 2D islands can be classified according to
the relative abundance of two types of facets, namely shallow and
steep facets only. The role of possible wetting films has also
been recently considered for the 3D system of InAs on GaAs(001)
\cite{wang}. However, in addition to assuming predefined shapes,
in all these studies the role of possible dislocations in the
islands has not been included.


In this work, we use an atomistic model to numerically study the
equilibrium shape of strained islands allowing for both elastic
and plastic strain relaxation without assumptions on predefined
shapes. We adopt a 2D model here, but extension to more realistic
3D systems is also possible. The reduced dimension allows to study
all possible configurations within feasible computer time. Most of
our results here are for relatively small, nanoscopic islands up
to a few hundred atoms in size to examine deviations from the
continuum theory of elasticity. In addition, we want to
investigate the role of dislocation nucleation in determining the
equilibrium shape and size of these islands. In an earlier
approach, elastic and plastic strain relaxation have also been
treated with a common formalism using vertically coupled
Frenkel-Kontorova layers of finite length \cite{zangwill}, where
only particle displacements parallel to the substrate are allowed.
However, such model does not provide a realistic description of
dislocation nucleation, whereas in our study, dislocations of
arbitrary type are allowed in the final equilibrium configuration.

In the 2D model we use for the strained adsorbate island and
substrate the atomic layers are confined to a plane. Interactions
between atoms in the system are described by a generalized
Lennard-Jones (LJ) pair potential \cite{zhen83} $U(r)$, modified
\cite{tru02a,tru02b} to ensure that the potential and its first
derivative vanish at a predetermined cutoff distance $r_c$, given
by
\begin{eqnarray}
&&U(r)=  V(r) , \qquad   r \leq r_0;  \cr &&U(r)=V(r) \left[ 3
\left( \frac {r_c-r}{r_c-r_0} \right) ^2 - 2 \left( \frac
{r_c-r}{r_c-r_0} \right) ^3 \right] , \ \  \ r > r_0, \cr &&
\label{LJ}
\end{eqnarray}
where
\begin{equation}
V_{\rm ab}(r)=  \varepsilon_{\rm ab} \left[ \frac m{n-m} \left(
\frac {r_0}r \right) ^n - \frac n{n-m} \left( \frac{r_0}r \right)
^m \right].
\end{equation}
Here, $r$ is the interatomic distance, $\varepsilon_{\rm ab}$ the
dissociation energy, which can be different for
substrate-substrate (${\rm ab}={\rm ss}$), adsorbate-adsorbate
(${\rm ab}={\rm ff}$) and adsorbate-substrate (${\rm ab}={\rm
fs}$) interactions, and $r_0$ is the equilibrium distance between
the atoms. For $n=6$ and $m=12$, $U(r)$ reduces to the standard
$6-12$ LJ potential with a smooth cutoff. For most of the
calculations we have chosen the values $n=5$ and $m=8$. In
contrast to the $6-12$ potential, this has a slower falloff. When
combined with the variation of the cutoff radius $r_c$, this
choice allows us to study the effect of the range of the
potential.  The equilibrium interatomic distance $r_0$ was set to
values $r_{\rm ss}=r_0$, $r_{\rm ff}$ and $r_{\rm fs}$ for the
substrate-substrate, adsorbate-adsorbate and adsorbate-substrate
interactions, respectively. The parameter $r_{\rm fs}$ for the
adsorbate-substrate interaction was simply set as the average of
the film and substrate lattice constants, {\it i.e.} $r_{\rm
fs}=(r_{\rm ff}+r_{\rm ss})/2$. The lattice misfit $f$ between the
adsorbate and the substrate can thus be defined as
\begin{equation}
f=(r_{\rm ff}-r_{\rm ss})/ r_{\rm ss}.
\end{equation}
A positive mismatch $f>0$ corresponds to compressive strain and
negative $f<0$ to tensile strain when the adsorbate island is
coherent with the substrate. Calculations were performed with
periodic boundary conditions for the substrate in the direction
parallel to the adsorbate-substrate interface. In the
calculations, the two bottom layers of the five-layer substrate
were held fixed to simulate a semi-infinite substrate while all
other layers were free to move. We checked that increasing the
substrate thickness did not affect the results. Typically, each
layer of the substrate contained about $100-500$ atoms.

To obtain the equilibrium shape of the island for a fixed total
number $N$ of atoms without assuming any predetermined shapes, we
use a systematic search approach. Each initial coherent
configuration is described by a set of integer numbers, ${n_i}$
specifying the number of atoms in successive layers of the island.
In terms of these quantities, the two types of facets, considered
in the previous works \cite{daruka99,daruka02} correspond to $n_i
- n_{i+1}=1$ for steep facets and $n_i - n_{i+1}=3$ for shallow
facets. The only physical restriction we impose is that the island
has a reflection symmetry about a line through the center and
overhangs are not allowed. Then, for each initial configuration,
molecular dynamics (MD) cooling was run to allow the system to
relax and reach a minimum energy configuration. The equilibrium
shape for a given $N$ is identified as the relaxed configuration
with lowest energy among all the possible configurations. In the
present case, this leads to complete relaxation of the interlayer
bonds in the islands, while intralayer bonds remain strained.

First, we present the results where the potential parameters were
chosen to be $\varepsilon_{\rm ss} = \varepsilon_{\rm ff} =
\varepsilon_{\rm fs} = 3410.1$ eV  and $r_0=2.5478$ {\AA}
corresponding to Cu substrate \cite{zhen83}. Phase diagrams for
the equilibrium shapes as a function of total number of island
atoms $N$ and lattice misfit $f$ are shown in Fig. 1(a), for two
different cutoff radii $r_c=3.82$ {\AA} and $r_c=5.3$ {\AA}, which
we call here the SR and LR cases, respectively. Different phases
are labelled by the total number of layers in the island. In
agreement with previous works \cite{politi00}, there is a
transition from a single layer configuration (uniform flat wetting
layer) to an island configuration, above a critical size or
lattice misfit. However, unlike the results from continuum
elasticity theory, the phase diagram is not symmetric with respect
to the misfit parameter and thus the behavior for compressive and
tensile strained layers is quite different. The asymmetry is more
pronounced for the LR interaction potential, which includes
significant contribution from next nearest neighbor atoms.

The predominant shapes of the islands in the phase diagram are
shown in Fig. 1(b). We performed a systematic classification of
these shapes for $1 \le N \le 105$ and $-10 \% \le f \le 10 \%$.
The strained film remains flat roughly between $-5 \% \le f \le 5
\%$ (SR) and $-3 \% \le f \le 7 \%$ (LR, up to $N=63$). The region
of 2D islands appears already for $N=11$ for the largest misfits,
and the extent of the flat film regime in the $f-N$ plane shrinks
for increasing $N$ (cf. Fig. 1(a)). The shapes A, C, and D
alternate with $N$, the shapes A and C being most common while B
is relatively rare.

The critical size along the transition line from single layer to
finite height island shape in our phase diagram can be fitted to a
power law as $N_c \propto f^{-a} $, with $ a \approx 3.8$,
consistent with the result from the continuum elasticity approach
\cite{daruka99}. Similar behavior has also been found in a model
of vertically coupled Frenkel-Kontorova chains \cite{zangwill}.
For the SR potential, a rough estimate through simple bond
counting yields a value for the ratio of the shallow to steep
facet surface energies $r \simeq 0.5$. According to Ref.
\onlinecite{daruka99}, this would imply an equilibrium shape of
only either the steep facet with truncated top or steep facet
followed by shallow facet and then truncated top, corresponding to
the shapes A and C in Fig. 1(b). Although these general features
agree with the elastic model calculations, there are important
differences. First, there is compressive-tensile asymmetry in the
shapes which is particularly pronounced for the LR potential. In
addition, for small islands (less than about $200$ atoms) there
are additional equilibrium shapes B and D in regions of the phase
diagram in Fig. 1(a) that were not accounted for in Ref.
\onlinecite{daruka99}. More recent work based on the elasticity
theory \cite{daruka02} finds that the possibility of shallow
facets below the steep ones, as observed in shape B, should also
be considered. However, shape D that we find here remains
unaccounted for by the continuum elasticity theory.


If the number of atoms $N$ is sufficiently large, then the optimal
size and shape for the island depend on  the subtle balance
between the strain energy, the surface energy of the island and
the interface energy between the island and the substrate.
Depending on the parameters of the interatomic potentials, phases
other than those shown in Fig. 1(a) can occur.
We consider here the situation where $\varepsilon_{\rm fs} >
\varepsilon_{\rm ss} = \varepsilon_{\rm ff}$.  For small misfit,
complete wetting occurs, as expected. However, at large enough
misfit {\it e.g.} $f=8$ \% and for $\varepsilon_{\rm
fs}/\varepsilon_{\rm ss} \approx 1.5$, the behavior is totally
different. In this case, for a system of lateral size $200$ and
for total number of atoms $N=600$, the minimum energy
configuration corresponds to that of a wetting layer with an
island of shape A with 26 layers high on top of it.  This result
also implies the  existence of an optimal size for a given shape
of the island. Our results thus correspond to the 2D version of
the 3D islands in Ref. \onlinecite{wang}. Thus by varying the
misfit and the substrate-adsorbate interactions, we find
configurations corresponding to the commonly known different modes
of adsorbate growth (Frank-Van der Merve, Stranski-Krastanow and
Volmer-Weber), which occur as minimum energy configurations in our
model. Detailed results for the different regimes will be
published elsewhere \cite{jalkanen05}.

We have also studied minimal energy paths for transitions from a
flat layer to an island of finite height. For a fixed number of
atoms $N$ and misfit $f$ close to the transition, we consider two
shapes corresponding to the different states across a transition
line from 1 to 2 ML in Fig. 1. Given these as initial and final
states, we use the Nudged Elastic Band (NEB) \cite{neb} method to
generate a (locally) minimum energy transition path between these
two different shapes. We follow the similar approach used in the
study of defect nucleation in strained epitaxial films, introduced
recently \cite{tru02b}. As an initial guess for the transition
path we use a simple linear interpolation. The resulting energy
profile and configurations along the minimum energy path are shown
in Fig. 2. As it is clear from the figure, there is a large energy
barrier separating the two distinct minima corresponding to the
two different shapes. This is generally true for any two states
bordering the transition line in our phase diagram in Fig. 1.
Thus, depending on the time scale, the transition may occur too
slowly to be observed during epitaxial growth. The existence of
this large energy barrier, obtained with our atomistic model,
supports the conclusion from the elastic theory calculations
\cite{daruka99,daruka02} that the transition can be regarded as
first order.

For sufficiently large islands or misfits we find that relaxing an
initial configuration with MD cooling already generates
dislocations in the lowest energy state. This implies that the
energy barrier for dislocation nucleation is zero or negligible
above a critical value. We find that the dislocations nucleate
from the edges of the adsorbate-substrate interface for an
initially dislocation-free island. This is in contrast with the
mechanism for a flat uniform film \cite{tru02b}, where
dislocations nucleate from the top layer. This result provides a
strong support from atomistic calculations for the conclusions
obtained within continuum elasticity theory
\cite{johnson,spencer01}. To better understand the dislocation
nucleation mechanism, we consider a region of phase space where
the dislocation is not necessarily spontaneously generated, {\it
i.e.}, there may exist a finite barrier for nucleation. The
transition path for dislocation nucleation is generated using the
NEB approach with the coherent island and the island with
dislocation as the initial and final states. As a initial guess
for the transition path we use again a simple linear interpolation
scheme between the coherent and incoherent states. The resulting
energy profile and configurations along the minimum energy path
found are shown in Fig. 3. The sequence of configurations along
the transition path shows that the two dislocations nucleate
successively at the separate edges of the island and then
propagate inwards. The energy barrier for a fixed island size
decreases with misfit. This is consistent with experimental
results \cite{chen}, which show that small islands are
dislocation-free but dislocations appear in the island when it
reaches a critical size.

In summary, we have studied the equilibrium shapes, shape
transitions and strain relaxation processes of strained,
nanoscopic epitaxial islands
The equilibrium shapes are determined via energy minimization for
different shapes without any preassumptions abut the possible
shapes, and they differ considerably from the elastic continuum
predictions. The  systematic transition path search combined with
the NEB approach  allow us to determine the energy barrier and
transition path for the shape transition and for dislocation
nucleation in initially coherent islands. In particular, we find
that dislocations can nucleate spontaneously at the edges of the
adsorbate-substrate interface above a critical size or misfit.
Although these calculations were performed for a 2D atomistic model,
the method can also be extended to more realistic 3D systems.

\section{Acknowledgements:} We wish to thank prof. Joachim Krug
for useful discussions. This work has been supported in part by
Funda\c c\~ao de Amparo \`a Pesquisa do Estado de S\~ao Paulo -
FAPESP (grant no. 03/00541-0) (E.G.) and the Academy of Finland
through its Center of Excellence program (T.A-N., J.J. and O.T.).

\renewcommand{\thefigure}{\arabic{figure}(\alph{subfig})}
\setcounter{subfig}{1}
\begin{wrapfigure}{l}{8cm}
\begin{minipage}[lt]{8cm}
\epsfig{file=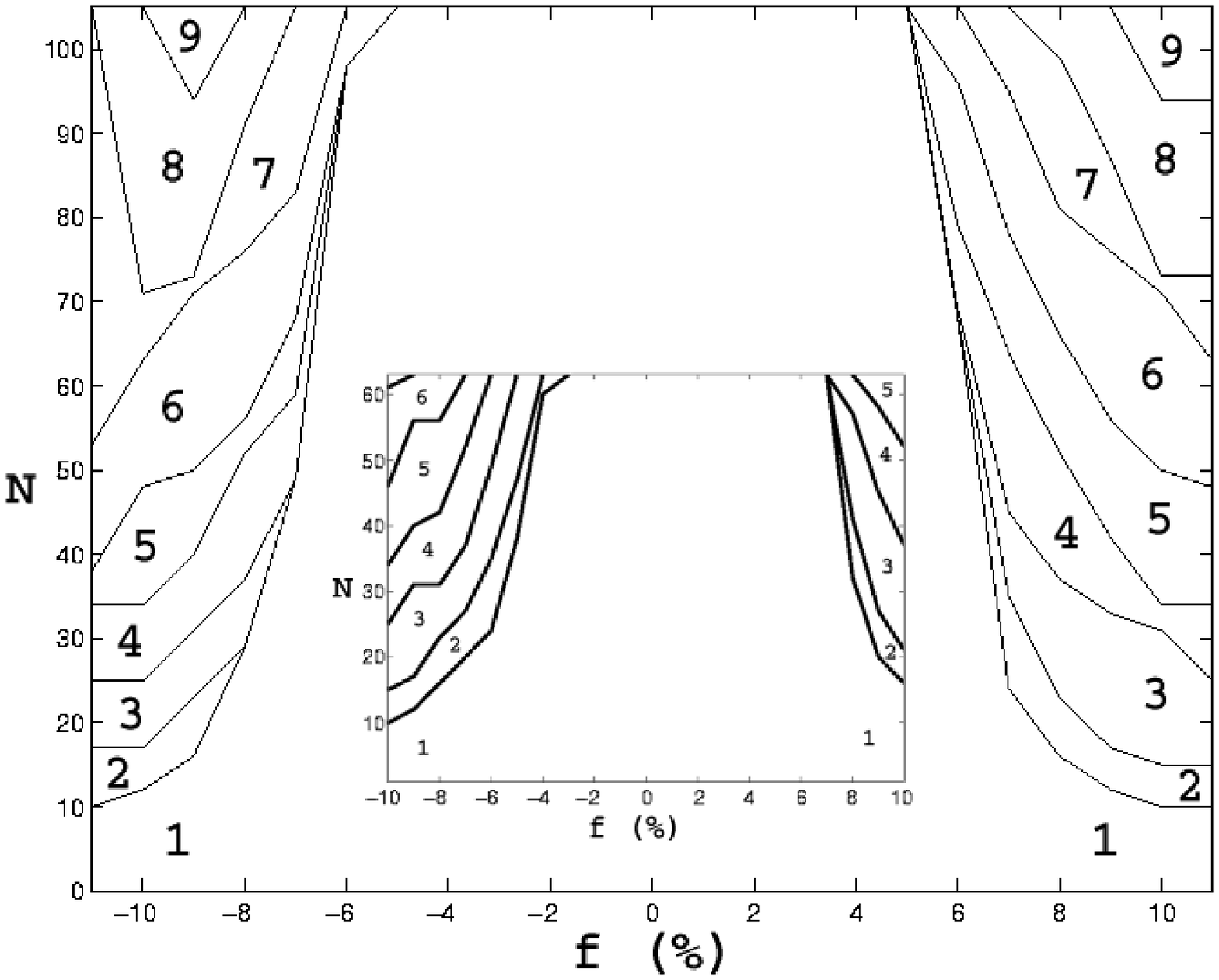,width=\textwidth}
\caption{Phase diagram showing island height as a function of
number of atoms $N$ and the lattice misfit $f$ for the SR (main
figure) and LR (inset) potentials. Different phases are labelled
by the number of layers in the island.}
\end{minipage}
\begin{minipage}[lb]{8cm}
\vspace{0.05\textheight}
\addtocounter{figure}{-1}
\addtocounter{subfig}{1}
\epsfig{file=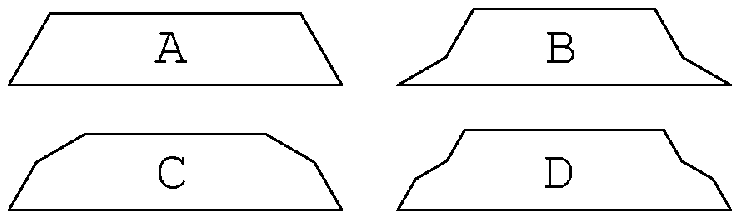,width=\textwidth}
\caption[a]{Predominant island shapes in the phase diagram for
nanoscopic islands. Shape D is not predicted by continuum
elasticity theory here.}
\end{minipage}
\end{wrapfigure}

\renewcommand{\thefigure}{\arabic{figure}}
\begin{wrapfigure}{r}{8cm}
\begin{minipage}[rt]{8cm}
\epsfig{file=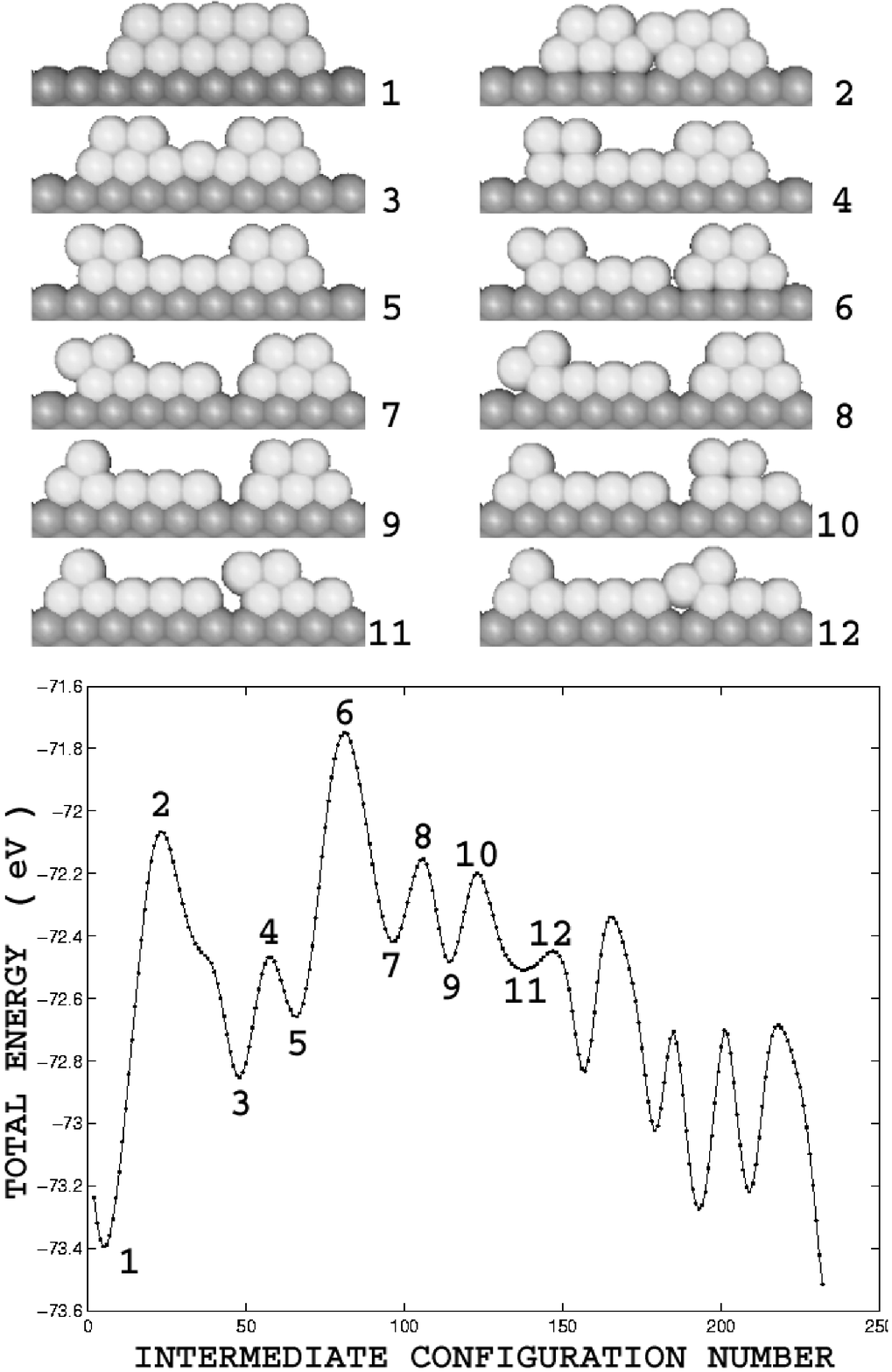,width=8cm,height=0.5\textheight}
\caption{First 12 particle configurations (upper frame)
corresponding to the labelled extrema in the energy profile (lower
frame). This minimum energy path has been generated by NEB
and it corresponds to the shape transition boundary for an 11-atom
island to a flat layer. The final flat configuration (not shown here) is
reached so that the remaining two second-layer atoms in
configuration 12 move close to the island edges and descend to the
substrate.}
\end{minipage}
\begin{minipage}[rb]{8cm}
\includegraphics[bb= 1cm  8cm  22cm   28cm, width= 8cm]{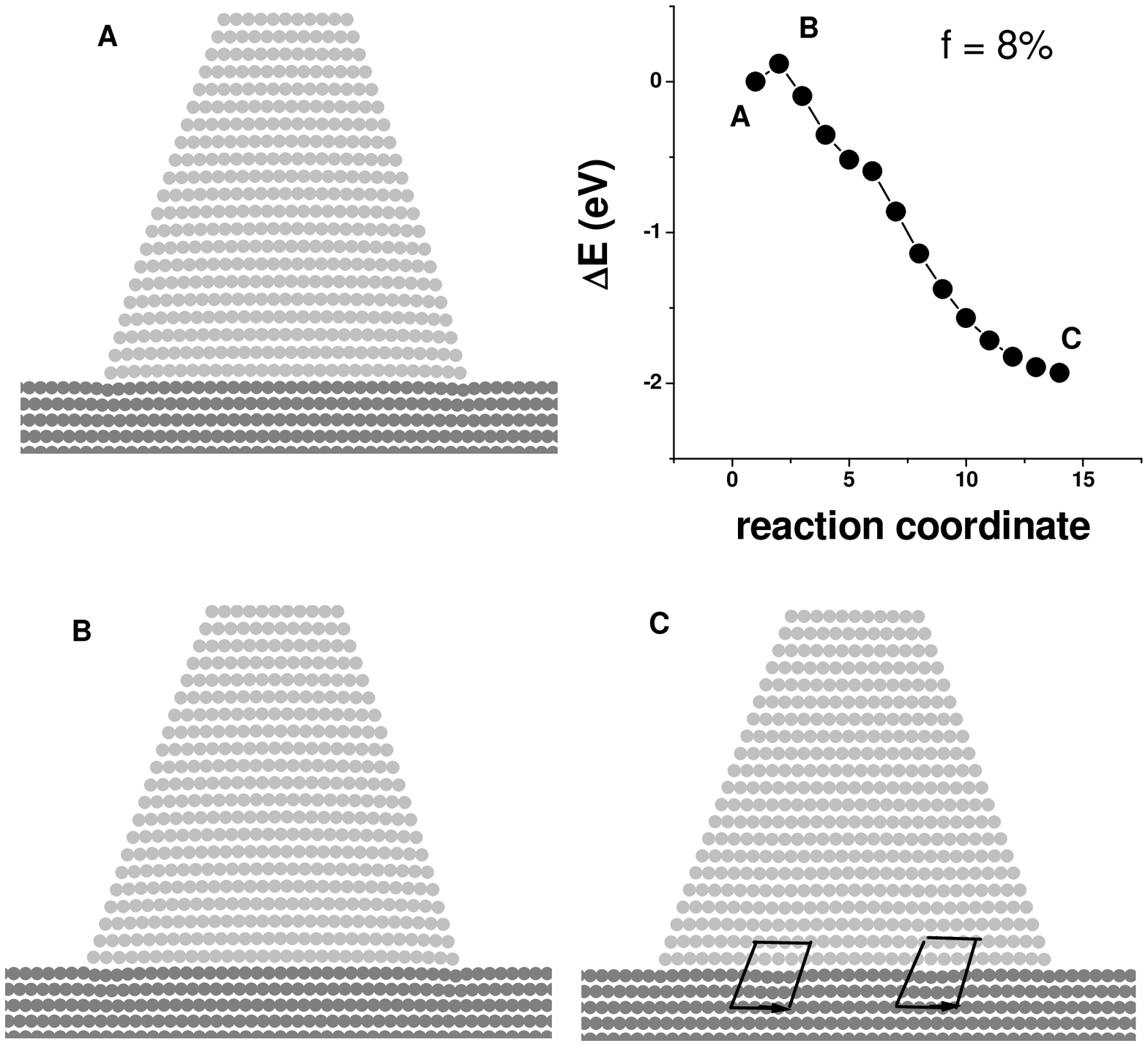}
\caption{NEB minimum energy profile and particle configurations
for a 411-atom strained island, where a pair of dislocations
nucleate at the edges of the island. Closed paths in (c) are the Burgers
circuits around the dislocation cores in the final state.
See text for details.}
\end{minipage}
\end{wrapfigure}

\end{document}